\documentclass{ws-procs975x65}

\usepackage{amsmath}
\usepackage{graphicx}
\usepackage{dcolumn}
\usepackage{bm}
\usepackage{amssymb}
\usepackage{latexsym}

\def\setC{\mathbb{C}}

\def\setN{\mathbb{N}}

\def\setR{\mathbb{R}}

\def\spose#1{\hbox to 0pt{#1\hss}}
\def\lta{\mathrel{\spose{\lower 3pt\hbox{$\mathchar"218$}}
     \raise 2.0pt\hbox{$\mathchar"13C$}}}
\def\gta{\mathrel{\spose{\lower 3pt\hbox{$\mathchar"218$}}
     \raise 2.0pt\hbox{$\mathchar"13E$}}}

\newcommand{\ie}{\textsl{i.e.~}}

\newcommand{\eg}{\textsl{e.g.~}}

\newcommand{\GReCO}{$\mathcal{G}\setR\varepsilon\setC\mathcal{O}$}

\newcommand{\Hu}{\mathcal{H}}
\newcommand{\Ka}{\mathcal{K}}

\begin{document}

\title{Propagating Cosmological Perturbations in a Bouncing Universe}

\author{Patrick Peter and J\'er\^ome Martin}

\address{Institut d'Astrophysique de
Paris, \GReCO, FRE 2435-CNRS, 98bis boulevard Arago, 75014 Paris,
France.\\
E-mails: peter@iap.fr, jmartin@iap.fr}

\maketitle

\abstracts{Using the simplest model for a bouncing universe, namely
that for which gravity is described by pure general relativity, the
spatial sections are positively curved and the matter content is a
single scalar field, we obtain the transition matrix relating
cosmological perturbation modes between the contracting and expanding
phases. We show that this case provides a specific example in which
this relation explicitely depends on the perturbation scale whenever
the null energy condition (NEC) is close to be violated.}

\begin{figure*}
\begin{center}
\includegraphics[width=9cm]{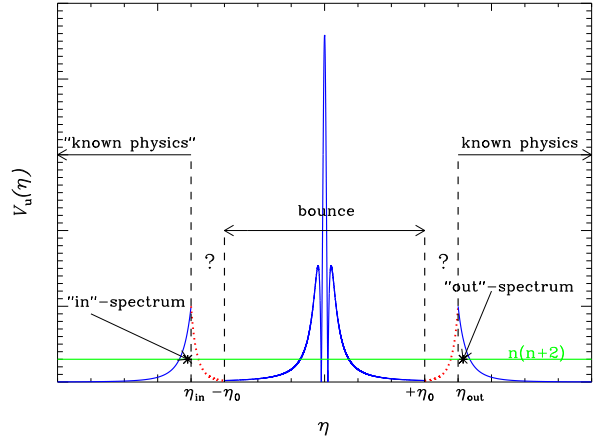}
\caption{The shape of the potential in our bounce model. The usual
assumption is that the potential starts as the full line growing curve
in the region indicated ``known physics'', then extends all the way to
the corresponding full line decaying curve in the other known physics
region with no special characteristics involved. In particular, the
potential is thus assumed to have a global maximum somewhere in
between, with perhaps a few irrelevant features on top of an otherwise
regular curve. This means that once the mode has crossed such a
potential, it will evolve through without noticeable change, except
perhaps in its amplitude, but no dependence in the wavenumber can be
introduced. By contrast, the actual potential we obtained in our model
must follow first the same growing full curve, then drop as the dotted
line in an essentially unknown way (hence the question mark), then
connect to the actual bounce part. This is followed by another largely
unknown piece connecting the really knwon physics. Clearly, the mode
intersects this potential many times, and as the bounce transition
itself changes the spectrum in a wavelength dependent way.}
\label{fig:potTOT}
\end{center}
\end{figure*}

Alternative models to inflation, many of which being possibly
examplified by the Pre Big-Bang (PBB) paradigm\cite{PBB}, propose a
phase preceeding the usual expanding (radiation and matter dominated)
epochs. In some instances, as \eg the Einstein frame description of
the PBB, this phase is a contracting phase which, in order to match
present time expansion observational data, demands a bounce to have
taken place. Although the contracting and expanding phases are in
principle controled by well known physics (quantum field theory in
curved spacetime), the bounce is often assumed to be describable by
high order terms in the underlying theory (superstring, say), and thus
not much can be said on this phase. The simplest possibility consists,
as discussed on Fig.~\ref{fig:potTOT}, in assuming the perturbation
modes of the expanding phase to be wavelength-independent linear
combinations of those in the contracting phase; this is in fact what
happens for long wavelengths in the inflation-to-matter domination
transition.

One way to perform a bounce, not necessarily taking into account the
perhaps essential ingredients of the abovementionned scenarios,
consists in using pure general relativity with a scalar field in a
potential, provided the spatial sections are positively
curved\cite{MPD}. In this case, a comoving wavelength is just an
eigenvalue of the Laplace operator on the 3-sphere, namely $k^2 =
n(n+2)$, with $n\in\setN$ (in practice, given the current constraints
on the curvature of space, the relevant modes are for $100\lesssim
n\lesssim 10^7$) and the corresponding perturbation equation of motion
can be cast in the form
\begin{equation}
\label{eomu}
u''+\left[n(n+2)-\frac{\theta ''}{\theta }-3\mathcal{K}(1-c_{_{\rm
S}}^2)\right]u \equiv u''+\left[ n(n+2)-V_u\left(\eta\right)
\right]u=0\, ,
\end{equation}
thereby defining the potential $V_u(\eta)$ as a function of the
conformal time $\eta$. The variable $u$ is related to the Bardeen
gravitational potential $\Phi$ through 
\begin{equation}
\Phi \equiv \frac{\kappa }{2}(\rho +p)^{1/2}u =\frac{\sqrt{3\kappa }}{2}
\frac{{\Hu}}{a^2\theta }u ,
\label{defu} 
\end{equation}
where the conformal Hubble factor $\Hu$ is defined by the scale factor
$a(\eta)$ through $\Hu\equiv a'/a$ and the function $\theta$ is
\begin{equation}
\label{deftheta}
\theta \equiv \frac{1}{a}\biggl(\frac{\rho }{\rho +p}\biggr)^{1/2}
\biggl(1-\frac{3{\Ka}}{\kappa \rho a^2}\biggr)^{1/2}=
\frac{1}{a}\biggl(\frac{3}{2\Gamma }\biggr)^{1/2}\, .
\end{equation}
while
\begin{equation}
c_{_{\rm S}}^2\equiv \frac{p'}{\rho'} = -\frac{1}{3} \left( 1+
2\frac{\varphi''}{\Hu \varphi'}\right).\label{cs2}
\end{equation}
stands for the ``sound velocity''.

When the bounce is approximated by the Taylor expansion (see
Ref.~\refcite{MPD} for the specific form of the coefficients), 
\begin{equation}
a(\eta )=a_0\biggl[1+\frac{1}{2}\biggl(\frac{\eta }{\eta _0}\biggr)^2
+\delta \biggl(\frac{\eta }{\eta _0}\biggr)^3+\frac{5}{24}(1+\xi )
\biggl(\frac{\eta }{\eta _0}\biggr)^4 \biggr]\, ,
\label{aseries}
\end{equation}
the potential $V_u$ in Eq.~(\ref{eomu}) takes the shape indicated at
the center of Fig.~\ref{fig:potTOT}, whose essential properties are
captured in the approximation $V_u(\eta) \simeq -C_\Upsilon
\delta(\eta)$, with $C_\Upsilon\propto (\eta_0^2-1)^{-1/2} \ll 1$, the 
last inequality holding when the bounce is the shortest non NEC
violating ($\eta_0$ close to unity). With such a $\delta$ potential,
Eq.~(\ref{eomu}), which is formally equivalent to a
``time''--independent Schr\"odinger equation, can be solved and give a
transition matrix relating the dominant $D_\pm$ and subdominant
$S_\pm$ modes of the contracting ($-$) and expanding ($+$) phases,
namely
\begin{equation}
\begin{pmatrix}
D_+\cr S_+
\end{pmatrix}
=
\begin{pmatrix} 
T_{11} & T_{12} \cr T_{21} & T_{22}
\end{pmatrix}
\begin{pmatrix}
D_- \cr S_-
\end{pmatrix}\, ,
\label{transT}
\end{equation}
and one finds\cite{MPD} that the transfer matrix leading behaviour
is
\begin{equation}\label{matrix}
T - \mathbf{1} \propto 
-\frac{i}{\sqrt{\Upsilon n(n+2)}}
\begin{pmatrix}
1&1\cr -1&-1
\end{pmatrix}
\end{equation}
\ie it contains a term proportional to the inverse of the wavenumber. Note
also that since none of the matrix element vanishes, one expects the
spectrum to exhibit specific properties that still need be
investigated in details.

It should be noted, by way of conclusions, that the matrix of
Eq.~(\ref{matrix}), obtained as a solution of a Sch\"odinger-like
equation, complies with the associated requirements of conservation of
``probabilities'' (namely that the reflection and transmission
coefficients sum up to unity), unitarity of the $S-$matrix that can be
built out of it and, in the symmetric $\delta=0$ case, of
time-reversal invariance\cite{prep}.

\section*{Acknowledgments}
We wish to thank Robert Brandenberger, Ruth Durrer and Fabio Finelli
for enlightening discussions.  

 \end{document}